\documentclass[reprint,
superscriptaddress,
amsmath,amssymb,
aps, 
pra,
twocolumn
]{revtex4-2}

\usepackage{graphicx}
\usepackage{physics,mathtools}

\usepackage{hyperref}
\hypersetup{%
   colorlinks = true,
   linkcolor=blue,
   citecolor=blue,
   urlcolor=blue
 }
 
\renewcommand{\Re}[1]{\text{Re}#1}
\renewcommand{\Im}[1]{\text{Im}#1}

\begin{document}
\title{Magic States in the Asymmetric Quantum Rabi Model}
\author{A. Campos-Uscanga}
\affiliation{Departamento de F\'isica, Universidad Aut\'onoma Metropolitana-Iztapalapa, Av. Ferrocarril San Rafael Atlixco 186, C.P. 09310 Mexico City, Mexico}
\author{E. Benítez Rodríguez}
\affiliation{Departamento de F\'isica, Universidad Aut\'onoma Metropolitana-Iztapalapa, Av. Ferrocarril San Rafael Atlixco 186, C.P. 09310 Mexico City, Mexico}
\author{E. Piceno Martínez}
\affiliation{Centro de Investigaciones en Óptica, Loma del Bosque 115, C.P. 37150, Guanajuato, Mexico}
\author{M. A.~Bastarrachea-Magnani}
\affiliation{Departamento de F\'isica, Universidad Aut\'onoma Metropolitana-Iztapalapa, Av. Ferrocarril San Rafael Atlixco 186, C.P. 09310 Mexico City, Mexico}
\email{bastarrachea@xanum.uam.mx}

%%%%%%%%%%%%%%%%%%%%%%%%%%%%%%%%%%%%%%%%%%%%%%%%%%
%%%%%%%%%%%%%%%%%%%%%%%%%%%%%%%%%%%%%%%%%%%%%%%%%%
\begin{abstract}
Non-stabilizerness is a resource for quantum computing that has been extensively studied in qudit networks. It describes the degree to which Clifford gates cannot generate a given state, capturing the advantage of quantum over classical computing. However, its definition in continuous variables and general composite systems remains an open issue. We study the magic resource in a bipartite system, the Asymmetric Quantum Rabi model, a paradigmatic model from quantum optics. We explore the presence of non-stabilizerness in the qubit-reduced system throughout the Hamiltonian parameter space, the role of light-matter interactions in its generation, and the manifestation of Wigner function negativity in the corresponding bosonic degree of freedom. Finally, we discuss our results for magic state preparation in the strong and ultra-strong coupling regimes within the context of quantum informational systems.
\end{abstract}
%%%%%%%%%%%%%%%%%%%%%%%%%%%%%%%%%%%%%%%%%%%%%%%%%%
%%%%%%%%%%%%%%%%%%%%%%%%%%%%%%%%%%%%%%%%%%%%%%%%%%


\maketitle

%%%%%%%%%%%%%%%%%%%%%%%%%%%%%%%%%%%%%%%%%%%%%%%%%%
%%%%%%%%%%%%%%%%%%%%%%%%%%%%%%%%%%%%%%%%%%%%%%%%%%
\section{Introduction}
\label{sec:1}
%%%%%%%%%%%%%%%%%%%%%%%%%%%%%%%%%%%%%%%%%%%%%%%%%%
%%%%%%%%%%%%%%%%%%%%%%%%%%%%%%%%%%%%%%%%%%%%%%%%%%

%%%%% Quantum supremacy
A goal of quantum computing is to attain quantum advantage, i.e., the idea that, by exploiting the unique properties of quantum systems, namely quantum correlations, a fault-tolerant quantum computer would outperform classical computers by either solving problems faster than they can, or solving those they are unable to~\cite{Feynman1982,Nielsen2011book}. There is an active search for benchmarking systems and protocols to achieve it~\cite{Proctor2022,Proctor2024}, where superconducting qubits and topological systems have set up a clear path~\cite{Arute2019}. A major challenge for quantum information has become identifying those features proper to quantum computing that would give it any advantage, particularly in speedup, to accomplish tasks within the scope of computational complexity~\cite{Preskill2012}. However, due to perturbations and decoherence, full access to arbitrary, scalable quantum operations is challenging to implement in physical systems. The problem then becomes determining the best way to perform quantum computation under operational restrictions dictated by physical constraints~\cite{Weedbrook2012,Veitch2014,Howard2017}.

%%%%% Resource Theory
In this context, {\it resource theories} turn necessary to characterize adequate resources for the optimal performance of quantum computing~\cite{Veitch2014,Coecke2016,Howard2017}. A resource theory is a framework that allows one to exploit physical phenomena for practical tasks~\cite{Chitambar2019,Horodecki2013}, in particular, it describes the elements that serve as the prime part to transmit information under different conditions, i.e., {\it resources}~ \cite{Bengtsson2017book}. Among the standard resources for quantum information, one finds quantum entanglement, quantum steering, and Bell non-locality~\cite{Wiseman2007, Bancal2014B,Piceno2023}. They have been applied in different protocols~\cite{Chitambar2019}, including quantum teleportation~\cite{Jin2010,Pirandola2015} and protocols for quantum key distribution~\cite{Pironio2009,Branciard2012,Xu2020}. 

%%%%% The concept of magic.
According to the Gottesman-Knill theorem~\cite{Gottesman1999,Aaronson2008}, quantum computers based only on stabilizer circuits can be classically simulated in polynomial time, so they have no computational advantage. While stabilizer states lead to stable fault-tolerant quantum computing~\cite{Gottesman1998}, the so-called {\it magic} or non-stabilizerness has been proposed as a resource to attain a speedup for universal quantum computing~\cite{Ahmadi2018}. Generally speaking, it amounts to the degree to which a quantum state cannot be generated by Clifford gates~\cite{Dai2022}. Mainly, non-stabilizerness has been investigated in the context of fault-tolerant quantum computation and its protocol of magic state distillation~\cite{Knill2004,Reichardt2005,Howard2017,Campbell2017,Souza2011}. Although a relatively new concept, it has been broadly studied in both the discrete~\cite{Veitch2014,Aaronson2008,Dai2022,Shuangshuang2023,Howard2017,Ahmadi2018} and continuous regimes~\cite{Ghose2007,Albarelli2018}, as well as in multipartite scenarios~\cite{Leone2022,Schwonnek2020,Cotfas2023,Turkeshi2023}. Additionally, there are current investigations about the relationship between non-stabilizerness and several fundamental topics such as  chaos~\cite{Leone2022,Turkeshi2023,Passarelli2024,Varikuti2025}, contextuality~\cite{Howard2014, Bermejo-Vega2017,Booth2022}, incompatibility~\cite{Ghai2023}, and many-body systems~\cite{Liu2022}.

%%%%% The problem of magic measurement. 
As it happened with entanglement before~\cite{Horodecki2009}, there is currently an ongoing discussion about quantifiers of {non-stabilizerness} that are both easy to calculate and sensitive enough to detect it. Many measures have been proposed including the sum negativity and mana~\cite{Veitch2014}, the (regularized) relative entropy~\cite{Piani2009,Veitch2014}, robustness~\cite{Howard2017,Heinrich2019}, stabilizer extent~\cite{Bravyi2016a,Bravyi2016b,Bravyi2019,Heimendahl2021}, thauma~\cite{Wang2020}, characteristic functions~\cite{Dai2022}, relative entropies~\cite{Liu2022}, R\'enyi entropies~\cite{Leone2022,Oliviero2022,Haug2023a,Leone2024}, the dyadic monotone~\cite{Beverland2020}, GPK~\cite{Hahn2022}, Bell measurements~\cite{Haug2023b}, and other monotones~\cite{Ahmadi2018,Warmuz2024}. Several of them are based on the negativity of the Wigner function~\cite{Gross2007,Veitch2014}, which works for qudit systems of prime odd dimension~\cite{Marchiolli2019,Schwonnek2020}. However, the system's dimension has become a limitation of the resource theory for non-stabilizerness, inherited from discrete phase space formulations~\cite{Gibbons2004,Wootters1987}.

%% The AQRM Hamiltonian.
The Quantum Rabi model (QRM)~\cite{Xie2017,LeBoite2017,Larson2017} is a paradigmatic model from quantum optics that accounts for a qubit strongly interacting with a quantum harmonic oscillator, typically a single-mode radiation field. Because the rotating-wave approximation (RWA) does not hold for larger light-matter coupling, the QRM is also used to describe the ultra-strong and deep-strong coupling regimes~\cite{FornDiaz2019,Kockum2019}. The biased, generalized, or asymmetric quantum Rabi Quantum model (AQRM)~\cite{Braak2011,Li2021a,Xie2021} is a modification that includes a bias term breaking the $\mathbb{Z}_2$ symmetry of the QRM. The AQRM has attracted interest recently because of the emergence of a hidden symmetry when the bias term equals a half-integer multiple of the bosonic frequency~\cite{Wakayama2017,Mangazeev2021,Li2021a,Batchelor2016,Liu2019,ReyesBustos2021,Xie2021,ReyesBustos2023}, now becoming a locus for the quantum integrability problem~\cite{Larson2013,Larson2017}. The QRM has found implementations in setups relevant to quantum technologies and quantum computing, including superconducting qubits~\cite{Blais2021} and trapped ions~\cite{Lv2018,Cai2021}, where the parameters become tunable. The bias term can be experimentally tuned in circuit QED via a qubit loop~\cite{Yoshihara2018}, in Cooper boxes coupled to optomechanical resonators~\cite{Armour2002,Irish2003}, and solid-state systems~\cite{Niemczyk2010,FornDiaz2010,Yoshihara2017}. 

%% Our contribution
As a description of spin-boson coupling, an essential feature of the AQRM is its intrinsic hybrid nature, which includes both discrete and continuous systems interacting with each other. The quantum properties of hybrid systems, such as quantum entanglement and non-locality, have been extensively studied in recent years~\cite{Costanzo2015,Jeong2014,Ma2020}, demonstrating that these systems can be directly applied in quantum technologies within the field of quantum computing. In this work, we explore the presence of magic states in the spectrum of the AQRM as a function of the spin-boson interaction and the anisotropy. To this end, we analyze the magic resource in both the traced qubit and photon subspaces for individual eigenstates of the AQRM and evaluate it within the parameter space. In the latter case, we employ negativity of the Wigner function as a resource for a continuous degree of freedom~\cite{Albarelli2018}. The dynamics of non-stabilizerness~\cite{Shuangshuang2022} and other measures of non-classicality~\cite{Dai2020,Fu2021} have been studied in the Rabi model under the RWA, i.e., the Jaynes-Cummings model~\cite{Jaynes1963}, where the time scales to attain maximum non-stabilizerness and oscillations were investigated for different initial states of light. 

Given that the USC regime has been proposed as a means to overcome several of the challenges for quantum computing, such as high-fidelity preparation of entangled states~\cite{Macri2018}, protecting a system against relaxation~\cite{Stassi2018}, reaching scalability via boosting qubit-qubit interactions~\cite{Stassi2020}, and creating ultra-fast quantum gates~\cite{Romero2012}, we believe the AQRM can be employed for the preparation of magic states via adiabatic evolution of the spectrum, similar to what has been done with entanglement~\cite{Joshi2016,Wang2018}, for both discrete and continuous degrees of freedom. But even more, our study of the magic resource in an interacting system upgrades non-stabilizerness from a quantum characteristic of individual systems to a way to witness how quantum correlations behave thanks to the interaction of the two quantum systems.

The article is organized as follows. In Sec.~\ref{sec:2}, we present the AQRM model and briefly discuss its spectrum and properties. In Sec.~\ref{sec:3}, we revisit the concept of non-stabilizerness, discuss its standard measures for the discrete Heisenberg-Weyl algebra, and their extensions to the continuous one. Next, in Sec.~\ref{sec:4}, we study {non-stabilizerness} over the energy spectrum of the Rabi model and its parameter space by considering the subsystems separately and identifying conditions for the emergence of magic resource. Finally, in Sec.~\ref{sec:6}, we discuss perspectives for magic state preparation and present our conclusions. We include several Appendices with details of the calculations.

%%%%%%%%%%
\begin{figure*}[ht]
    \includegraphics[width=1.0\textwidth]{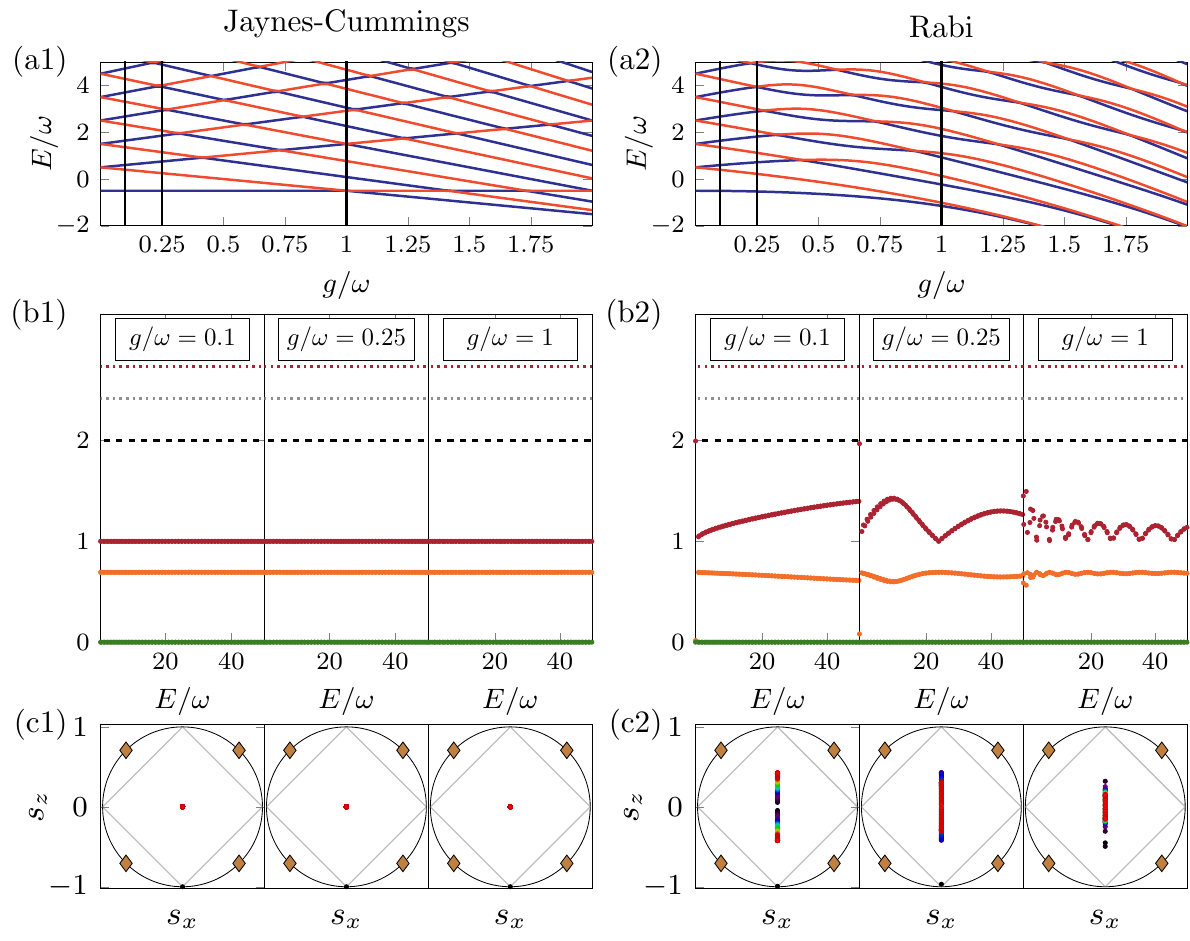}
    \caption{(Top) Spectrum of the AQRM for (a1) $\xi=0$ (Jaynes-Cummings limit) and (a2) $\xi=1$, in resonance $2\omega=\Delta$ as a function of the spin-boson interaction, $g$, for $\epsilon/\omega=0$. Vertical lines indicate selected values of the spin-boson coupling $g/\omega=0.1$, $0.25$, and $0.5$. (Middle) Witness and monotone of non-stabilizerness for (b1) $\xi=0$ and (b2) $\xi=1$. $\mathcal{M}$ (red points), $\text{mana}$ (green points), and von Neumann entropy (orange points) of the atomic reduced density matrix of individual eigenstates, for $\epsilon/\omega=0$, and the corresponding coupling values indicated in the top row ($g/\omega=0.1,0.25,1$). The witness $\mathcal{M}$ determines the presence of non-stabilizerness when it takes values above 2, indicated by the black dashed line. Gray and red dotted lines signal the value of $\mathcal{M}$ of the $\vert H\rangle$ and $\vert T\rangle$ states, respectively. (Bottom) We show the qubit reduced density matrix for each eigenstate in the Bloch sphere from the corresponding case in the middle row, for (c1) $\xi=0$ and (c2) $\xi=1$. The lowest (highest) energy states are in blue (red). Diamonds signal the $|H\rangle$ states.}
  \label{fig:1}
\end{figure*}
%%%%%%%%%%

%%%%%%%%%%%%%%%%%%%%%%%%%%%%%%%%%%%%%%%%%%%%%%%%%%
%%%%%%%%%%%%%%%%%%%%%%%%%%%%%%%%%%%%%%%%%%%%%%%%%%
\section{Asymmetric Quantum Rabi model}
\label{sec:2}
%%%%%%%%%%%%%%%%%%%%%%%%%%%%%%%%%%%%%%%%%%%%%%%%%%
%%%%%%%%%%%%%%%%%%%%%%%%%%%%%%%%%%%%%%%%%%%%%%%%%%
The asymmetric quantum Rabi model (ARQM) describes the interaction between a single radiation mode and a biased two-level system~\cite{Klockl2016} (tending to the $x$-direction). The Hamiltonian is
\begin{gather} 
\hat{H}_{\xi}=\omega\hat{a}^{\dagger}\hat{a}+\Delta\hat{Z}+\label{eq:AQRM_Ham}\\ \nonumber
g\left[(1+\xi)\left(\hat{a}+\hat{a}^{\dagger}\right)\hat{X}+(1-\xi)\left(\hat{a}-\hat{a}^{\dagger}\right)i\hat{Y}\right]+\epsilon\hat{X},
\end{gather}
where $\omega$ and $2\Delta$ are the frequencies of the boson and qubit, respectively, $g$ is the Rabi splitting, $\hat{a}$ ($\hat{a}^{\dagger}$) is the boson annihilation (creation) operator and $\hat{X}$, $\hat{Y}$, and $\hat{Z}$ are the Pauli spin matrices; $\epsilon$ is the asymmetry or bias parameter. $\xi\in[0,1]$ is an anisotropy parameter that creates a family of AQRM Hamiltonians, allowing for a continuous tune from the Jaynes-Cummings ($\xi=0$)~\cite{Jaynes1963} to the Rabi ($\xi=1$)~\cite{Rabi1936,Rabi1937} limits.

For $\epsilon=0$, in the Jaynes-Cummings limit, the Hamiltonian becomes integrable in the classical sense due to the conservation of the total number of excitations operator $\hat{\Lambda}=\hat{a}^{\dagger}\hat{a}+\hat{Z}+\frac{1}{2}\hat{\mathbb{I}}$, generating an $U(1)$ symmetry. In contrast, in the Rabi limit, the parity operator $\hat{\Pi}=\exp(i\pi\hat{\Lambda})$ becomes conserved instead, so a $\mathbb{Z}_{2}$ symmetry appears~\cite{Larson2007,Braak2019}. Still, its spectrum can be solved by taking advantage of the existing symmetry, $\mathbb{Z}_{2}$, in the Bargmann space representation~\cite{Braak2011,Braak2013,Braak2015}, or via Bogoliubov operators~\cite{Chen2012}. This sparked a discussion about the integrability of the QRM, and several approaches have been employed to address the issue~\cite{Yu2012}. The spectra of the standard Jaynes-Cummings and Rabi models as a function of the light-matter coupling are shown in Fig.~\ref{fig:1}. 

Solving the asymmetric model ($\epsilon\neq 0$) using a similar method is possible. Despite that the bias term breaks the remaining discrete $\mathbb{Z}_{2}$ symmetry, it was conjectured that a hidden symmetry appears~\cite{Li2015,Wakayama2017,Ashhab2020,Li2021a,Lu2022}. This is a non-conventional, parameter-dependent symmetry that manifests explicitly for fixed values of the bias, $2\epsilon=n\omega$ with $n$ integer~\cite{Li2021a}, where energy levels exhibit crossings related to a conic intersection in the energy landscape~\cite{Batchelor2016,Li2021b}. These crossings are referred to as Juddian exceptional points~\cite{Judd1979} and can be expressed in operator form as a series expansion of $\epsilon$~\cite{Mangazeev2021,ReyesBustos2021}.

In addition, the anisotropy $\xi$ leads to the anisotropic or generalized Rabi model~\cite{Xie2014,Tomka2014,Liu2017,Ying2022}, another variant of the QRM sharing features related to solvability. For any value of $\xi\neq 0$, the $U(1)$ symmetry is broken and becomes a $\mathbb{Z}_{2}$ symmetry. However, the balance between the rotating and counter-rotating terms causes the eigenstates belonging to different parity sectors to swap places in pairs~\cite {Xie2014}, also leading to a richer phase diagram~\cite{Ying2022}.

Our interest in the asymmetric model stems from the bias, as it provides a controllable means to rotate the qubit in the Bloch sphere and, alongside the light-matter interaction strength, a way to induce non-stabilizerness in the qubit subsystem, and as we will delve in the following to control the amount of Wigner function negativity of the bosonic subsystem.

%%%%%%%%%%%%%%%%%%%%%%%%%%%%%%%%%%%%%%%%%%%%%%%%%%
%%%%%%%%%%%%%%%%%%%%%%%%%%%%%%%%%%%%%%%%%%%%%%%%%%
\section{Measures of non-stabilizerness}
\label{sec:3}
%%%%%%%%%%%%%%%%%%%%%%%%%%%%%%%%%%%%%%%%%%%%%%%%%%
%%%%%%%%%%%%%%%%%%%%%%%%%%%%%%%%%%%%%%%%%%%%%%%%%%
A resource theory allows the organization of a scenario in terms of allowed or free and forbidden operations~\cite{Chitambar2019}. Free operations can be implemented without expending resources in a context inspired by physical constraints, generating free states. On the contrary, a forbidden operation requires something additional, i.e., a resource, to be performed. In this language, stabilizer operations and stabilizer states correspond to the free operations and free states, respectively~\cite{Veitch2014,Albarelli2018}. In quantum computation, we are interested in the stability of a circuit, that is, its robustness against noise. Then, we can make a quantum computing protocol stable by codifying it with stabilizer circuits. Every stabilizer circuit is made of gates from the Clifford group (Hadamard, Phase, and CNOT gates) and is meant to stabilize a fault-tolerant quantum computation. 

Due to the Gottesman-Knill theorem, efficient, universal quantum computing would require performing operations outside the Clifford group. Hence, non-stabilizer or magic states become a resource~\cite{Bravyi2005}. The protocol, thus, implies a process to generate magic states from Clifford gates. Simply put, a state possesses the magic resource if it is not a stabilizer state. For qubit systems, the magic states lie in the exterior of an octahedron whose vertices are the stabilizer states, and maximum non-stabilizerness would correspond to those states generated by the $\hat{H}$ and $\hat{T}$ matrices. There are eight magic states of the type $|T\rangle$ and twelve of the type $|H\rangle$, for instance, two of them are
\begin{gather}
|H\rangle\langle H|=\frac{1}{2}\left[\hat{\mathbb{I}}+\frac{1}{\sqrt{2}}\left(\hat{\sigma}_{x}+\hat{\sigma}_{z}\right)\right],\\
|T\rangle\langle T|=\frac{1}{2}\left[\hat{\mathbb{I}}+\frac{1}{\sqrt{3}}\left(\hat{\sigma}_{x}+\hat{\sigma}_{y}+\hat{\sigma}_{z}\right)\right].
\end{gather}
where the $|H\rangle$ states lie in the planes formed by the canonical directions in the Bloch sphere, particularly, the plane $x-z$. 

To quantify the magic resource in a quantum state $\hat{\rho}$, it is desirable to employ a monotone $Q(\hat{\rho})$ fulfilling two essential requirements: it vanishes over stabilizer states, and, under free operations, it is monotonically decreasing, representing their effortless character~\cite{Horodecki2013,Chitambar2019}. Finding good quantifiers of non-stabilizerness represents a challenge because one needs them to be easily computable and sensitive enough to the resource. That is why witnesses, i.e., dichotomic indicators that signal the presence of non-stabilizerness, are also employed~\cite{Horodecki1996,Dai2022}.  

Additionally, there is a distinction between discrete (qudit) systems and systems with continuous degrees of freedom in the context of a resource theory.  Such differences, also present in other quantum resources such as entanglement and steering, are crucial in describing, detecting, and measuring non-stabilizerness with appropriate witnesses and monotones~\cite{Veitch2014,Dai2022,Leone2024,Albarelli2018,Chabaud2024}. Here, we focus on one monotone: {\it mana}, i.e., the normalized negativity of the Wigner function, and the Dai-Fu-Luo characteristic function witness,~$\mathcal{M}$~\cite{Dai2022}, based on characteristic functions. Mana offers an operational meaning to the negativity of the Wigner function, so it is possible to extend it to continuous degrees of freedom and thus describe a resource for the bosonic subsystem in the AQRM.

%%%%%%%%%%%%%%%%%%%%%%%%%%%%%%%%%%%%%%%%%%%%%%%%%%
\subsection{Magic states of the qubit subsystem}
%%%%%%%%%%%%%%%%%%%%%%%%%%%%%%%%%%%%%%%%%%%%%%%%%%

For a qudit system, described by a $d$-dimensional Hilbert space $\mathcal{H}_{d}$, non-stabilizerness is often measured by exploiting the properties of the discrete Heisenberg-Weyl group~\cite{Appleby2005}. Consider $\mathbb{Z}_{d}={0,1,..,d-1}$ the modular ring of dimension $d$, the qudit states in the computational basis are $|j\rangle$, with $j\in\mathbb{Z}_{d}$. The set of operations over the qudit is given in terms of the generalized Pauli matrices acting in the computational basis~\cite{Schwinger1960}, the  phase operator
\begin{gather}\label{eq:z}
\hat{Z}=\sum_{j=0}^{d-1}\tau^{2j}|j\rangle\langle j|,
\end{gather}
and the shift operators
\begin{gather}\label{eq:x}
\hat{X}=\sum_{j=0}^{d-1}|j+1\rangle\langle j|,\,\,\,\hat{Y}=\sum_{j=0}^{d-1}|j-1\rangle\langle j|,
\end{gather}
with $\tau=-e^{i\pi/d}$. To construct a discrete phase space representation in the $\mathbb{Z}_{d}\otimes\mathbb{Z}_{d}$ space, one employs the unitary Heisenberg-Weyl or discrete displacement operators~\cite{Zhu2016}
\begin{gather}\label{eq:d}
\hat{D}_{k,l}=\tau^{kl}\hat{X}^{k}\hat{Z}^{l},
\end{gather}
The set $\mathcal{P}_{d}=\left\{\tau^{j}\hat{D}_{kl}|j,k,l\in\mathbb{Z}_{d}\right\}$ is the discrete Heisenberg-Weyl group. 

For arbitrary $d$, the computational basis can be transformed into the standard spin basis with spin $s=d/2+1$. See App.~\ref{app:1} for details about this. Moreover, for odd prime dimensions $d$, the discrete parity operator $\hat{P}_d$ is given by 
\begin{gather}
\hat{P}_{d}=\frac{1}{d}\sum_{l,k\in\mathbb{Z}_{d}}\hat{D}_{kl},
\label{eq:parityOP}
\end{gather}
allowing for the discrete phase-space representation of the qudit system in terms of the discrete Wigner function~\cite{Wootters1987,Gibbons2004,Bjork2008}
\begin{gather}\label{eq:diswig}
W^{d}_{k,l}(\hat{\rho})=\text{tr}\left(\hat{\rho}\hat{D}_{kl}\hat{P}_{d}\hat{D}_{lk}^{\dagger}\right)=\text{tr}\left(\hat{\rho}\hat{A}^{d}_{k,l}\right),
\end{gather}
where $\hat{A}^{d}_{k,l}=\hat{D}_{kl}\hat{P}_{d}\hat{D}_{kl}^{\dagger}$ are the phase point operators (PPOs)~\cite{Wootters1987}. It should be noted that the discrete Wigner function has no universal definition for all $d$, except for odd prime power dimensions~\cite{Dai2022}, where it has been shown that it is the only quasi-distribution covariant under Clifford operations~\cite{Gross2006,Gross2007}. However, the Wigner function can be defined with different constraints not necessarily related to the dimension of the systems ~\cite{Klimov2010, Ruzzi2005, Marchiolli2019, Rundle2021,Tilma2011,Tilma2016}.

For dimension 2, given the properties of the Pauli spin matrices, there is no unique form for the Wigner function~\cite{Wootters1987,Rundle2021}. Here, we employ the \textit{bona fide} discrete Wigner operator proposed in~\cite{Marchiolli2019}
\begin{gather}\label{eq:wigMAr}
\hat{W}^{(2)}_{k,l}=\frac{1}{2}\left[\mathbb{I}+(-1)^{l}\hat{X}+(-1)^{k+l+1}\hat{Y}+(-1)^{k}\hat{Z}\right].
\end{gather}
Note that the discrete Wigner function in Eq.~(\ref{eq:wigMAr}) and that proposed in~\cite{Wootters1987} are equivalent for our case because all the states the qubit can be found in are confined to the $x-z$ plane in the Bloch sphere (see App.~\ref{app:DiscreteWigner}). 

Here, we study the mana and Dai-Fu-Luo's characteristic function for $d=2$. The mana is a monotone that exploits the idea that the negative part of the Wigner function can be used as a resource~\cite{Veitch2012}. It reads~\cite{Veitch2014}
\begin{gather}
\text{mana}(\hat{\rho})=\log_{2}(2\text{sn}(\hat{\rho})+1),
\end{gather}
where 
\begin{gather}
2\text{sn}(\hat{\rho})+1 = \sum_{k,l\in\mathbb{Z}_d}|W_{\hat{\rho}}(k,l)|
\end{gather}
with $\text{sn}(\hat{\rho})$ the sum negativity~\cite{Veitch2014}, i.e., the sum of all the negative parts of the discrete Wigner function in Eq.~\ref{eq:diswig}. The mana has the advantage of being additive in the product of density matrices, and offers bounds to distillation protocols~\cite{Veitch2014}. On the other hand, the Dai-Fu-Luo function $\mathcal{M}$, that is the Fourier transform of the characteristic function in the discrete space, is defined as~\cite{Dai2022} 
\begin{gather}
\mathcal{M}(\hat{\rho})=\sum_{k,l\in\mathbb{Z}_{2}}
\left| \text{tr}\left(\hat{\rho}\hat{D}_{k,l}\right)\right|
\label{eq:DaiFuLuo}
\end{gather}
The advantages of this quantity are that it is well-defined for any dimension, it is easy to compute, its kernel is a simplified form of the Wigner function, and, as explained later, it is related to the von Neumann entropy. 

%%%%%%%%%%%%%%%%%%%%%%%%%%%%%%%%%%%%%%%%%%%%%%%%%%
\subsection{Magic states for the bosonic system}
%%%%%%%%%%%%%%%%%%%%%%%%%%%%%%%%%%%%%%%%%%%%%%%%%%

The bosonic part of the system is described by the Heisenberg-Weyl algebra $\text{hw}(4)$~\cite{Loebl1975}. Then, an extrapolation of the concept of the magic resource in the boson-reduced system is algebraically possible. We follow the idea of using Wigner function negativity as a resource for continuous variable operations, established in~\cite{Albarelli2018}. In this context, free states correspond to states with a completely positive Wigner function; those operations preserving Wigner positivity become free operations, and states with negativity in their Wigner function play the role of magic states~\cite{Albarelli2018}. The formulation is generalized in the standard phase space of canonical position and momentum $(q,p)$, where coherent states, $\hat{a}|\alpha\rangle=\alpha|\alpha\rangle$, are the most straightforward states for a continuous degree of freedom with a completely positive Wigner function. Notice that states with negativity in their Wigner function present explicit quantum behavior such as Fock states, cat states, states resulting from individual photon subtraction or addition, and cubic-phase states~\cite{Lutkenhaus1995,Gottesman2001,Kenfack2004,Biswas2007}. It is worth highlighting that, when considering the limit when the dimension of the discrete system tends to infinity $d\rightarrow\infty$, the discrete Heisenberg-Weyl algebra tends to the continuous one, giving us, in this manner, a relation between discrete and continuous phase spaces and, as a byproduct, a connection between discrete and continuous magic resource theories, as discussed in App.~\ref{app:1}. 

As usual, coherent states can be obtained by applying a (continuous) displacement operator
\begin{gather} \label{eq:dinf}
    \hat{D}(\alpha) = \exp\Big(\alpha\hat{a}^{\dagger}-\bar{\alpha}\hat{a}\Big),
\end{gather}
such that $\ket{\alpha} = \hat{D}(\alpha)|0\rangle_{f}$, with $|0\rangle_{f}$ the bosonic vacuum state. Correspondingly, the standard Wigner function is written as~\cite{Wigner1932}
\begin{gather} \label{eq:wig}
   W_{\hat{\rho}}(q,p) = \frac{1}{2\pi}\int\dd{q'}\mel{q-\frac{\hbar q'}{2}}{\hat{\rho}}{q+\frac{\hbar q'}{2}}e^{-ipq'}.
\end{gather}
Another way to express it is through the Wigner operator, $\hat{W}_\alpha$,~\cite{Ben-Benjamin2016}
\begin{gather}
W_{\hat{\rho}}(\alpha)=tr\left(\hat{\rho}\hat{W}_{\alpha}\right)=\text{tr}\left(\hat{\rho}\hat{D}(\alpha)\hat{P}\hat{D}(\alpha)^{\dagger}\right),
\end{gather}
where $\hat{P}$ is the parity operator, defined as $\hat{P}=\int\dd{x}\ketbra{-x}{x}$ in configuration space. The Hudson theorem, which establishes that pure states with positive Wigner function have a Gaussian behavior~\cite{Hudson1974,Soto1983}, sets up the idea that mana can be employed as a monotone, just like in the discrete case. The counterpart of mana is what is called Wigner logarithmic negativity~\cite{Albarelli2018}, which we name here as bosonic mana and reads, in terms of the Wigner function, as
\begin{gather}
\text{mana}_{\text{bos}}(\hat{\rho}_{B})=\log_{2}\left[\int \frac{d^{2}\alpha}{\pi}\abs{W_{\alpha}(\hat{\rho})}\right].\label{eq:WLN}
\end{gather}

%%%%%%%%%%
%%%%%%%%%%
\begin{figure}[ht]
      {\includegraphics[width=0.9\linewidth]{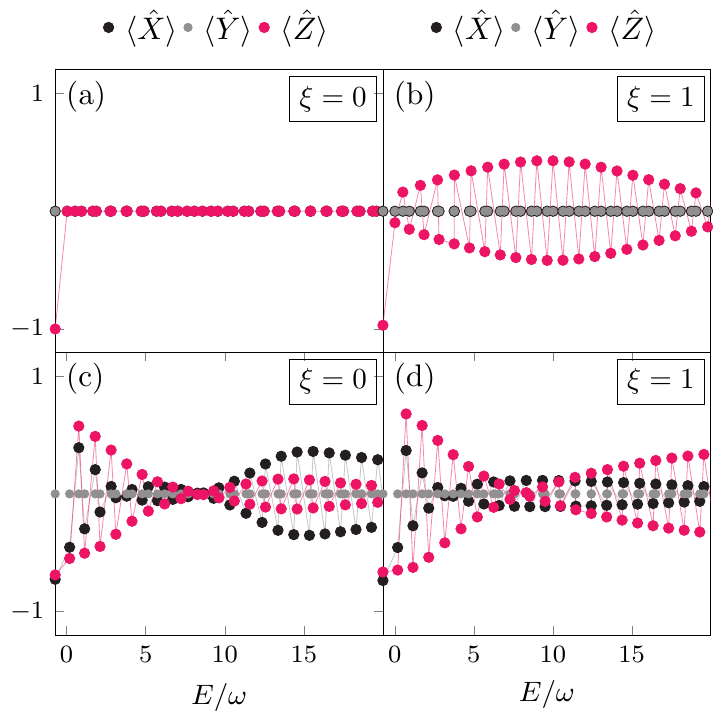}}
  \caption{Expectation values of $\hat{X}$, $\hat{Y}$ and $\hat{Z}$ as functions of energy $E/\omega$, for (top) $\epsilon=0$ and (bottom) $\epsilon=0.5\omega$ and (left) the Jaynes-Cummings ($\xi=0$) and (right) the Rabi ($\xi=1.0$) limits. $\langle\hat{Z}\rangle$ oscillates for all cases, but $\langle\hat{X}\rangle$ only in presence of the bias.} 
 \label{fig:2Z}  
\end{figure}
%%%%%%%%%%
%%%%%%%%%%

%%%%%%%%%%
\begin{figure*}[ht]
   \includegraphics[width=1.0\textwidth]{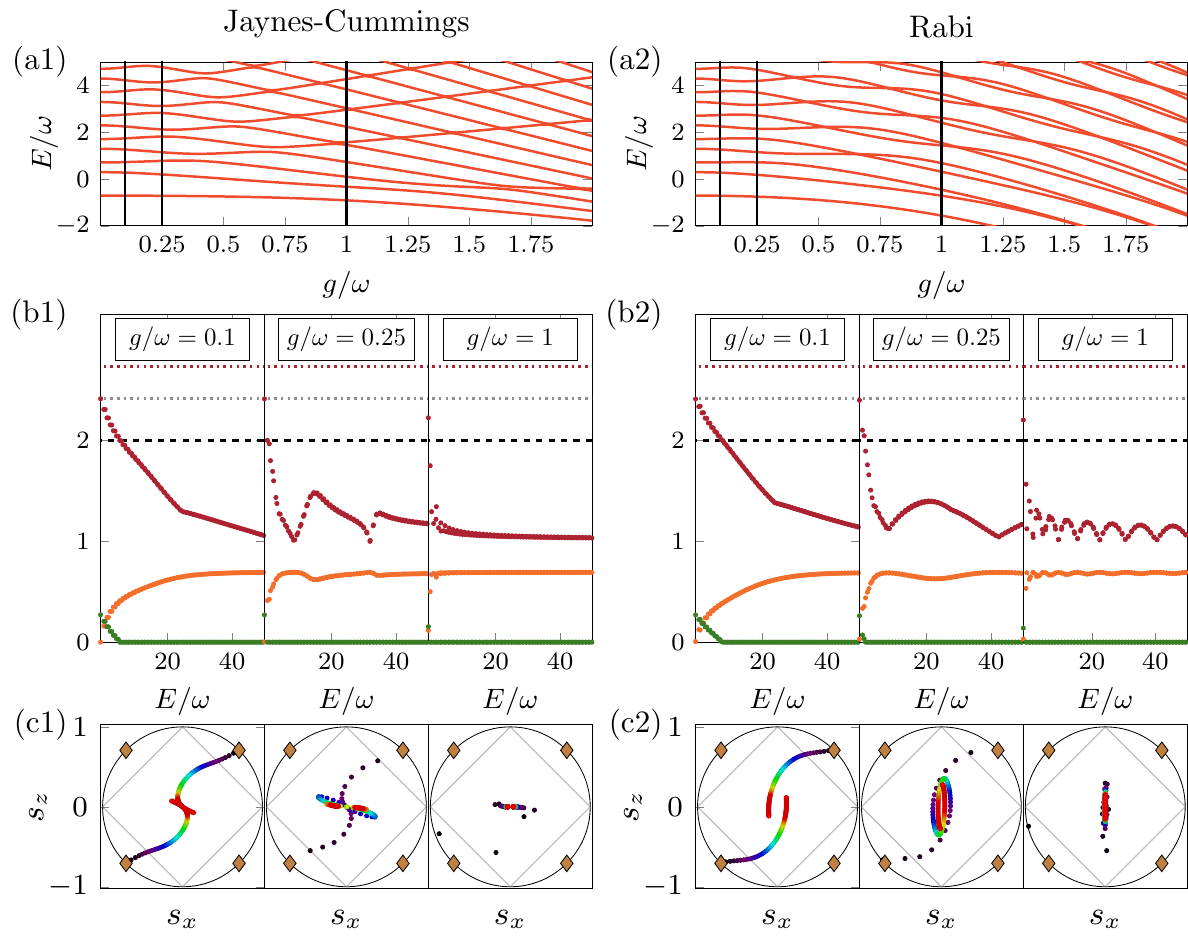}
    \caption{Same as in Fig.~\ref{fig:1} but for $\epsilon=0.5\omega$. Notice that the eigenstates are no longer identified by parity.}
    \label{fig:3}
\end{figure*}
%%%%%%%%%%
%%%%%%%%%%
\begin{figure*}[ht]
   \includegraphics[width=1.0\textwidth]{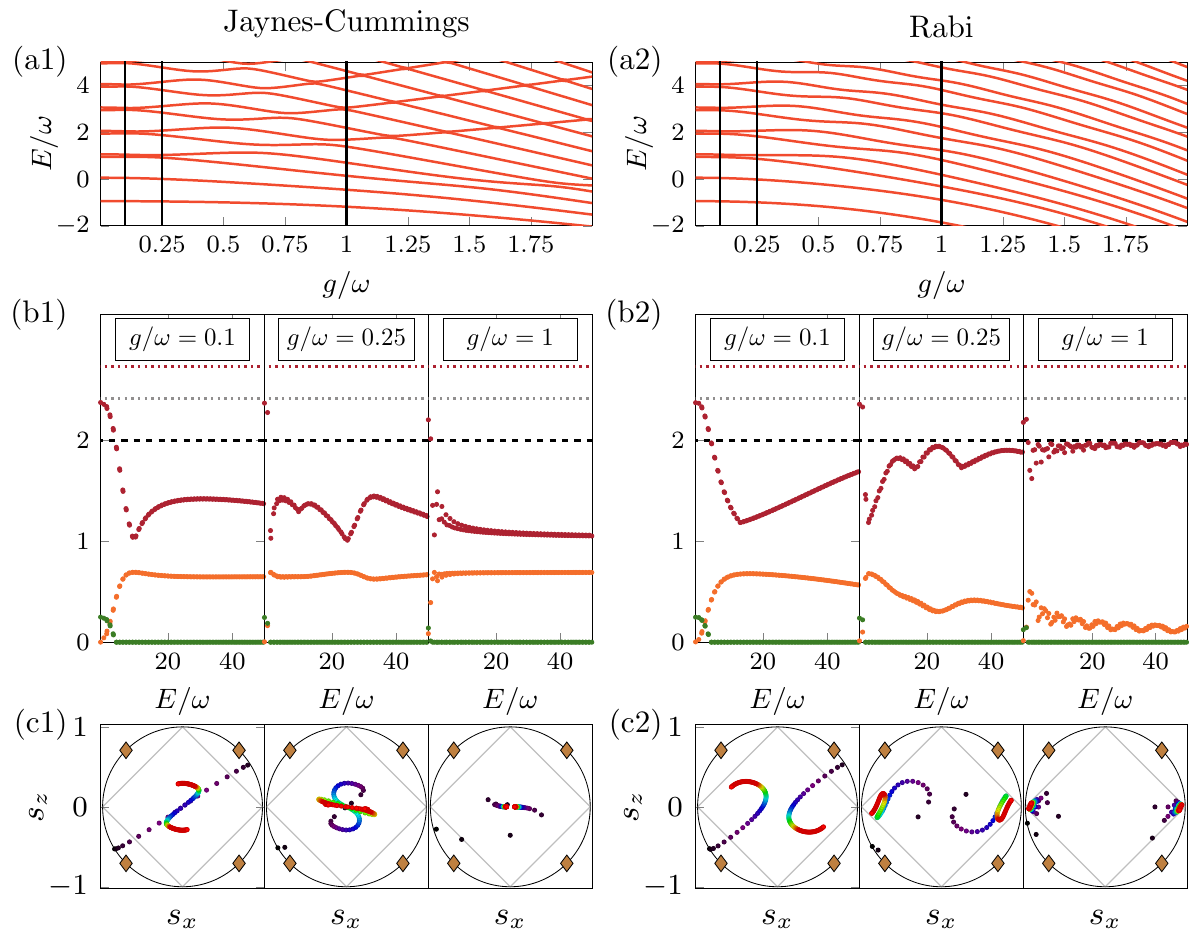}
    \caption{Same as in Fig.~\ref{fig:1} but for $\epsilon=0.8\omega$. Notice that the eigenstates are no longer identified by parity.} 
  \label{fig:4}
\end{figure*}
%%%%%%%%%%

%%%%%%%%%%%%%%%%%%%%%%%%%%%%%%%%%%%%%%%%%%%%%%%%%%
%%%%%%%%%%%%%%%%%%%%%%%%%%%%%%%%%%%%%%%%%%%%%%%%%%
\section{Magic resource in the spectrum of the AQRM}
\label{sec:4}
%%%%%%%%%%%%%%%%%%%%%%%%%%%%%%%%%%%%%%%%%%%%%%%%%%
%%%%%%%%%%%%%%%%%%%%%%%%%%%%%%%%%%%%%%%%%%%%%%%%%%

We solve numerically the AQRM Hamiltonian for a given set of parameters and truncate the boson space (up to $n_{\text{max}}$ bosons) such that the wave functions obey a numerical convergence criterion given in App.~\ref{app:numericalmethod}. Next, we study non-stabilizerness in individual eigenstates of $\ket{E_k}$ with a pure density matrix $\hat{\rho}_{k} =\ketbra{E_{k}}$, by tracing out either the boson $\hat{\rho}_{k,\text{S}}= \tr_{\text{B}}\left(\hat{\rho}_{k}\right)$ or the qubit $\hat{\rho}_{k,\text{B}} = \tr_{\text{S}}\left(\hat{\rho}_{k}\right)$ degree of freedom to focus on the resources in that particular subspace.

%%%%%%%%%%%%%%%%%%%%%%%%%%%%%%%%%%%%%%%%%%%%%%%%%%
\subsection{Qubit magic resource in the spectrum without asymmetry}
%%%%%%%%%%%%%%%%%%%%%%%%%%%%%%%%%%%%%%%%%%%%%%%%%%

We first explore the behavior of the $\text{mana}$ and Dai-Fu-Luo $\mathcal{M}$ for the atomic reduced density matrix of individual eigenstates for the JC ($\xi=0$) and standard QRM ($\xi=1$) models in the absence of asymmetry ($\epsilon=0$) and as a function of the light-matter coupling in the resonant case, i.e., when $\omega=2\Delta$. Also, we calculate the von Neumann entropy in the qubit subspace, i.e., $S_{\text{S}}(\hat{\rho}_{k,\text{S}})=\text{Tr}(\hat{\rho}_{k,\text{S}}\log\hat{\rho}_{k,\text{S}})$, which gives us a measure of the purity of the traced system. In this case, we know there is no magic resource in the eigenstates, because the expectation value of the $\hat{X}$ is zero over the eigenstates given parity conservation in both the JC and the QRM, so the reduced system is inside the octahedron in the Bloch sphere. Moreover, in all the cases considered in this work, the expectation value of $\hat{Y}$ is always zero. 

The spectrum of the JC and QRM Hamiltonians is shown in Figs.~\ref{fig:1} (a1) and (a2), respectively, where eigenstates belonging to a given parity are depicted in a different color (blue for even, red for odd parities). Next, we study our quantities of interest for three representative values of the light-matter coupling, from the perturbative ($g/\omega=0.1$) to the strong ($g/\omega=0.25$), and USC domains ($g/\omega=1.0$)~\cite{Rossato2017}. We recall that the JC model is no longer valid for larger couplings~\cite{Jaynes1963}. Therefore, for large $g/\omega$, this is merely a parametric exploration of the resource.

In Figs.~\ref{fig:1} (b1) we show the $\text{mana}$ (green points),  the $\mathcal{M}$ (red points), and von Neumann entropy (orange points) as a function of energy for the JC model for the three selected values of the light-matter coupling. In this case, we observe that both the $\text{mana}$ is exactly zero and $\mathcal{M}=1$ for all eigenstates and values of $g/\omega$, given that the quantity $\mathcal{M}$ indicates the presence of non-stabilizerness only when it takes a value greater than 2, thus, confirming the absence of the resource. Here, however, the constant behavior is a result of reduced system properties, as the von Neumann entropy becomes $S_{\text{S}}=\ln(2)$. The reason is that, in resonance, the qubit reduced matrix of the eigenstates of the JC Hamiltonian becomes maximally mixed. This can be explained easily in terms of the mixing angle $\theta_{\Lambda}$, defined as $\tan(\theta_{\Lambda})=2g\sqrt{\Lambda}/(2\Delta-\omega)$, which is contained in the coefficients of the eigenstates belonging to the number of excitations $\Lambda>0$ subspaces. In the resonant case, $\theta_{\Lambda}=\pi/2$ regardless of the values of $\Lambda$ and the light-matter coupling $g$. This is confirmed in Fig.~\ref{fig:1} (c1), where we show the location of the reduced state of each eigenstate in the Bloch sphere, by calculating $s_{z}=\text{tr}(\hat{\rho}_{\text{S}}\hat{Z})$ and $s_{x}=\text{tr}(\hat{\rho}_{\text{S}}\hat{X})$. We notice that the resonant case is very particular. From the equation above, out of resonance, the behavior of the JC case is similar to that of the QRM, as it will become evident later when we discuss the asymmetric cases.

The situation is different in the QRM. In the weak coupling regime, the ground state is close to separability. Hence, the von Neumann entropy is almost zero, and the $\mathcal{M}$ indicates a stabilizer state $\mathcal{M}\simeq 2$, as seen in Fig.~\ref{fig:1} (b2). Then, as one goes higher in energy, the eigenstate rapidly becomes a light-matter mixture, and a monotonic increase (decrease) of $\mathcal{M}$ ($S_{\text{S}}$) emerges. Nevertheless, again $\mathcal{M}\leq 2$ and the $\text{mana}$, as it is identical to zero, indicates the absence of the magic resource. This is verified geometrically by observing the location of all reduced density matrices in the Bloch sphere in Fig.~\ref{fig:1} (c2), where all states have $s_{x}=0$ and are inside the octahedron (excepting the ground-state).

Remarkably, the entropy and the characteristic function $\mathcal{M}$ are anticorrelated. We can explain this because for $d=2$, a functional connection between these two quantities can be obtained taking into consideration the effect of the symmetries $U(1)$ of the Jaynes-Cummings model and the $\mathbb{Z}_2$ of the Rabi model, both via $\langle\hat{X}\rangle=\langle\hat{Y}\rangle=0$. As discussed in App.~\ref{app:conjecture}, this leads to
\begin{gather}
\mathcal{M}(\hat{\rho})=1-\frac{dS}{d\rho}+O(\rho^{3}).
\end{gather}
Thus, the decrease of the $\mathcal{M}$ is accompanied by an increase in the von Neumann entropy. We will observe this behavior independently of the coupling and bias value for both models.

Increasing the coupling introduces a series of oscillations in the $\mathcal{M}$ (and von Neumann entropy) as a function of energy, as seen in Fig.~\ref{fig:1} (b2). We attribute this to the onset of manifolds of specific values for the total number of excitations in the individual eigenstates. As one traverses the spectrum, the lowest energy of that manifold becomes an instantaneous ground state in the parametric level evolution, becoming more separable than neighboring states~\cite{Kloc2017}.
This behavior cannot be seen from the Bloch sphere, as shown in Fig.~\ref{fig:1} (c2), but it is associated with oscillations in $\langle\hat{Z}\rangle$. In Fig.~\ref{fig:2Z} (a) and (b), we show this observable as a function of energy for the Jaynes-Cummings and QRM, respectively. The absence or presence of oscillatory behavior corresponds to what is observed in both the von Neumann entropy and $\mathcal{M}$ in the corresponding Figs.~\ref{fig:1} (b1) and (b2). The $\text{mana}$ is insensitive to these oscillations.

To introduce non-stabilizerness in the system, we need the bias term, so the expectation value of $\langle\hat{X}\rangle$ can become different from zero. 

%%%%%%%%%%%%%%%%%%%%%%%%%%%%%%%%%%%%%%%%%%%%%%%%%%
\subsection{Qubit magic resource in the spectrum with asymmetry}
%%%%%%%%%%%%%%%%%%%%%%%%%%%%%%%%%%%%%%%%%%%%%%%%%%

Next, we incorporate the asymmetry into the models. In Figs.~\ref{fig:3} (a1)-(a2) and~\ref{fig:4} (a1)-(a2), we show the spectra for two selected values of $\epsilon=0.5\omega$ and $\epsilon=0.8\omega$, respectively. Since, in general, the standard parity symmetry $\mathbb{Z}_{2}$ is now broken, we use a single color throughout the entire spectrum. 

As it can be observed from Fig.~\ref{fig:3} (b1)-(b2), the behavior of the $\text{mana}$, the $\mathcal{M}$, and, correspondingly, of the von Neumann entropy is quite similar between the asymmetric JC and Rabi models. For weak coupling $g/\omega=0.1$ (where the two models behave similarly), the ground state is close to a pure state, but its reduced state is not a stabilizer one anymore. The $\text{mana}$ becomes different from zero, and the $\mathcal{M}$ goes almost to the $1+\sqrt{2}$, strikingly, the value corresponding to the $|H\rangle$ states, indicating the presence of non-stabilizerness. As anticipated, the asymmetry breaks the parity symmetry, resulting in a rotation in the Bloch sphere. The resource emerges in the low-energy spectrum, where entropy also indicates a purer state. Then, mixedness develops as energy increases, and non-stabilizerness is lost. This can be understood in terms of the Bloch sphere, as shown in Figs.~\ref{fig:3} (c1) and (c2), where we observe that the ground-state is outside the octahedron and very close to the $|H\rangle$ state marked with a diamond. Next, the states go inside the octahedron, and the resource is lost. 

As for $\epsilon=0$, while the $\text{mana}$ only has information of magic states, the $\mathcal{M}$ and von Neumann entropy exhibit oscillations, as shown in Figs.~\ref{fig:3} (b1) and (b2). For $\epsilon=0.5\omega$, the main difference between the JC and Rabi models is that the oscillatory behavior becomes more pronounced as the coupling increases toward the USC regime. In Fig.~\ref{fig:2Z} (c) and (d), we show the expectation values of phase and shift operators as a function of energy, for the JC and Rabi models. Unlike the case of $\epsilon=0$, now both $\langle\hat{X}\rangle$ and $\langle\hat{Z}\rangle$ oscillate (and $\langle\hat{Y}\rangle$ remains zero). Still, in both cases, the entropy and $\mathcal{M}$ tend on average to the value of the most mixed state. This is reflected in the Bloch sphere, where we observe the states encircling toward the center of the sphere, as shown in Figs.~\ref{fig:3} (c1) and (c2).

Notoriously, it seems light-matter interactions play a role in suppressing non-stabilizerness. Figs.~\ref{fig:3} (b1) and (c1) show that, for both models, increasing $g/\omega$ reduces the number of states possessing the resource, until in the USC regime only the ground-state remains. Moreover, the value of $\mathcal{M}$ and the $\text{mana}$ of the ground state decrease as well. Therefore, there is a competition between the light-matter interaction and the bias to generate magic resource in the ground state and the lower part of the spectrum.

One would naively think that further increasing the asymmetry will always favor the presence of non-stabilizerness; however, as seen in Figs.~\ref{fig:4} (b2), where we study the case of the QRM at $\epsilon=0.8\omega$, the entropy decreases and the $\mathcal{M}$ approaches the value for stabilizer states as we increase $g/\omega$. This can be understood from the Bloch sphere in Fig.~\ref{fig:4} (c2), as all the eigenstates concentrate over the $x$-axis, i.e., toward another stabilizer state. In fact, for large values of $g/\omega$, one can perform the so-called polaron transformation~\cite{Lee2012,Jaako2016,DeBernardis2018,Pilar2020} $\hat{U}=\exp\left[-(g/\omega)\hat{X}\left(\hat{a}^{\dagger}-\hat{a}\right)\right]$, that shifts the bosonic operators as $\hat{A}=\hat{a}+(g/\omega)\hat{X}$. So, in the polaron frame, the AQRM reads (with $\xi=1$)
\begin{gather}
\hat{H}_{1}^{'}=\hat{U}^{\dagger}\hat{H}_{1}\hat{U}=\omega\left[ \hat{A}^{\dagger}\hat{A}-(g/\omega)^{2}\mathbb{I}+(\epsilon/\omega)\hat{X}\right]+\Delta\hat{Z}
\end{gather}
The $\hat{A}^{\dagger}$ ($\hat{A}$) creation (annihilation) operator describes a light-matter excitation in the strongly correlated regime and becomes exact when $\Delta/g\ll 1$ and $\epsilon=0$. We observe that if the asymmetry becomes dominant in the USC, one tends to the stabilizer states from the $\hat{X}$ operator. This makes clearer how strong-light matter interactions play against the generation of magic states. 

We aim to identify regions where the ground state exhibits maximum non-stabilizerness across the parameter space. Before that, we want to study what happens to the boson-reduced system within a resource theory based on the negativity of the Wigner function, i.e., a bosonic magic resource.

%%%%%%%%%%%%%%%%%%%%%%%%%%%%%%%%%%%%%%%%%%%%%%%%%%
\subsection{Bosonic magic resource over the spectrum}
%%%%%%%%%%%%%%%%%%%%%%%%%%%%%%%%%%%%%%%%%%%%%%%%%%

Because we have a hybrid system, one wonders what happens to the resources contained in the other reduced system: the bosons. Using the spectrum of the AQRM, we get the Wigner logarithmic negativity of the traced qubit subsystem using a numerical method based on expressing our states in the Fock basis~\cite{Curtrigth2001} (see App.~\ref{app:numericalmethod}). Unlike the qubit system case, we notice that for $g=0$, the Wigner logarithmic negativity is different from zero, because the eigenstates are Fock states $|n\rangle$. Therefore, our reference for the calculation of the Wigner logarithmic negativity is given by the Fock states, whose Wigner function is~\cite{Kenfack2004}
\begin{gather}
	W_{\ketbra{n}}(q,p) = \frac{(-1)^n}{\pi}\exp[-(q^2+p^2)]\mathcal{L}_n[2(q^2+p^2)],
\end{gather}
where $\mathcal{L}_n$ denotes the $n$-th order Laguerre polynomial. It must be emphasized that the Wigner function of the coherent state $\ket{n=0}$ is a Gaussian function centered at the origin and corresponds to the most classical state in the context of continuous degrees of freedom described above. As such, its Wigner negativity is zero. While the ground state lacks the resource, for $n\neq 0$ we will have it, even in the absence of light-matter interaction.

\begin{figure*}[ht]
    \includegraphics[width=1.25\columnwidth]{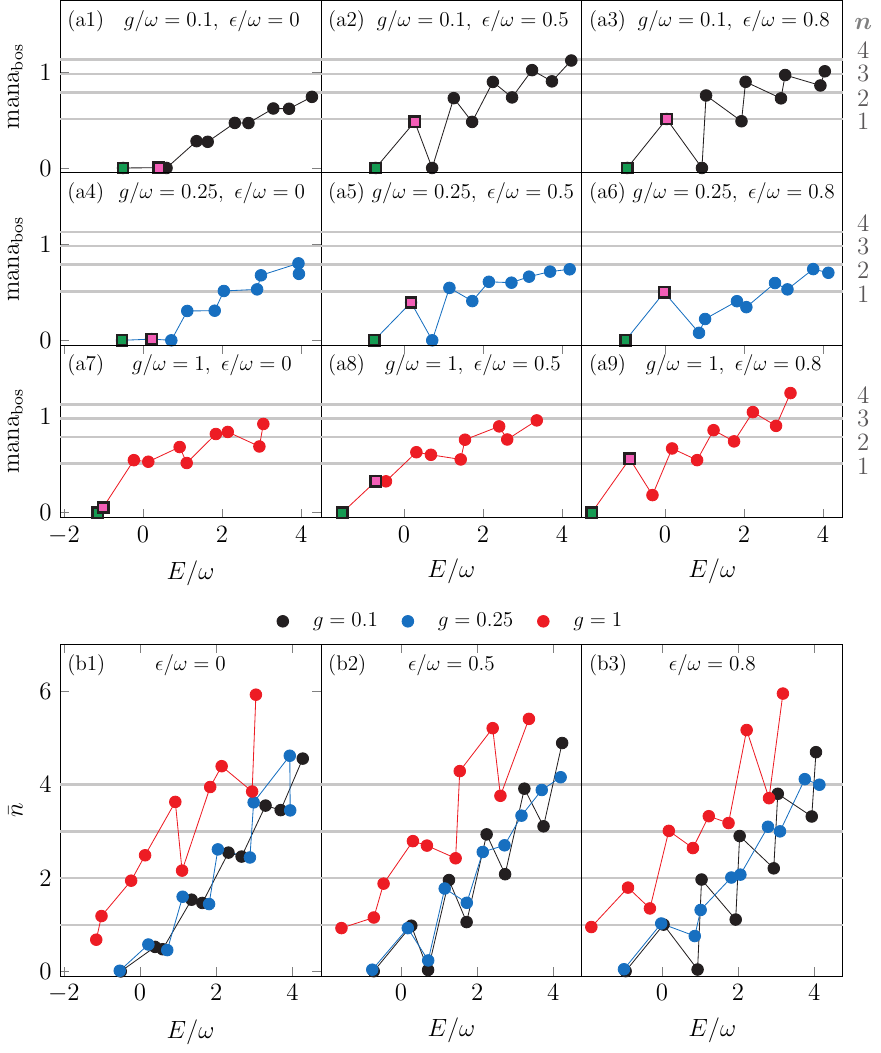}
    \caption{(a1)-(a9) Wigner logarithmic negativity or bosonic mana for the bosonic reduced density matrix of individual eigenstates of the AQRM, for a selected combination of parameters corresponding to those in Figs~\ref{fig:1},~\ref{fig:3} and ~\ref{fig:4}, as indicated, with $\omega=2\Delta=1$. We mark the ground and first excited states with green and pink squares, respectively. The gray horizontal lines indicate the negativity of Fock states with $ n=0, 1, 2, 3$, and $4$. (b1)-(b3) Mean number of bosons as a function of energy for weak ($g/\omega=0.1$), strong ($g/\omega=0.25$), and ultra-strong ($g/\omega=1$) couplings for three values of asymmetry, the oscillatory behavior is more pronounced as the asymmetry increases.} 
  \label{fig:5}
\end{figure*}

The results are shown in Fig.~\ref{fig:5} (a1)-(a9) for the same set of parameters in previous Figs.~\ref{fig:1},~\ref{fig:3}, and~\ref{fig:4} as indicated in the figure, and for the low-lying spectra. For all the parameter sets, we observe that the ground state indeed has zero bosonic mana, which will be denoted by $\text{mana}_{\text{bos}}$, corresponding to a state with zero photons.

The behavior of the bosonic mana as one increases energy does depend on the Hamiltonian parameters. Increasing the coupling does not change significantly the values, as they appear more closely related to the boson number described by the reduced state. This is verified in Figs.~\ref{fig:5} (b1)-(b3), where we plot the corresponding expectation value of the boson number $\bar{n}=\langle\hat{a}^{\dagger}\hat{a}\rangle$ in each eigenstate. The bosonic mana and mean number of bosons have a very similar behavior, a fact that makes sense because the negativity of the Wigner function is an increasing function of the number of bosons due to the intrinsic negativity in Fock states. For larger values of the asymmetry, there are oscillations in mana with a stair-like behavior, related to the breaking of the parity symmetry that mixes manifolds with different excitations. 

In summary, the bosonic degree of freedom contains a magic, continuous variable resource directly related to the average number of photons expressed by the eigenstate. It seems independent of the results on the qubit side.

%%%%%%%%%%
%%%%%%%%%%

%%%%%%%%%%
\begin{figure}[ht]
    \includegraphics[width=1\columnwidth]{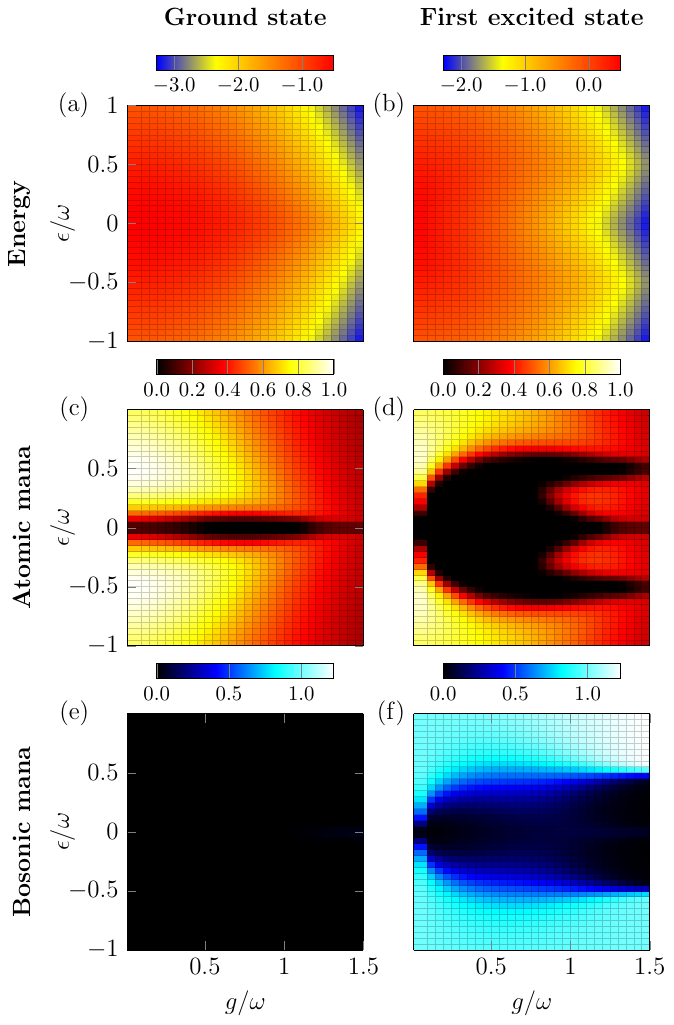}
    \caption{Non-stabilizerness for the ground- and first-excited states of the AQRM as a function of the bias $\epsilon$ and coupling $g$, for $\xi=1.0$. (Top) The energy of the ground (a) and the first-excited states (b). (Middle) Mana of the qubit-reduced ground (c) and the first excited states (d). (Bottom) Negativity of the boson-reduced ground (e) and the first excited states (f). Here, $\xi=1$, $\omega=2\Delta=1$.}
  \label{fig:6}
\end{figure}
%%%%%%%%%%

%%%%%%%%%%
\begin{figure}[ht]
    \includegraphics[width=1\columnwidth]{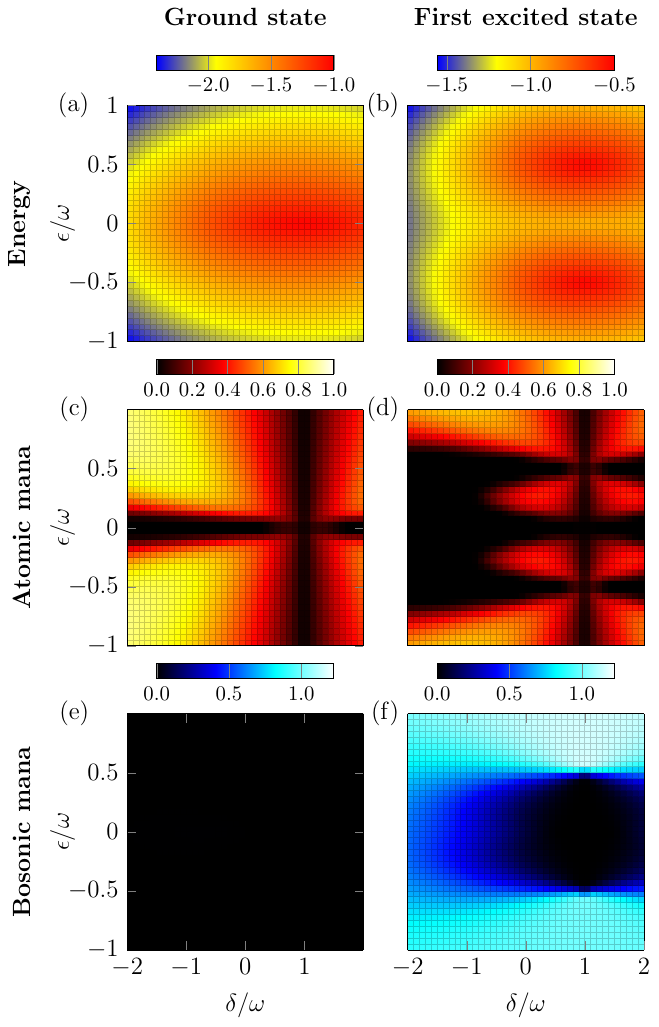}
    \caption{Same as in Fig.~\ref{fig:6}, but as a function of the bias $\epsilon$ and spin-boson detuning $\delta=\omega-2\Delta$, $\xi=1$, and $g/\omega=1$ (the USC).} 
  \label{fig:7}
\end{figure}
%%%%%%%%%%

%%%%%%%%%%%%%%%%%%%%%%%%%%%%%%%%%%%%%%%%%%%%%%%%%%
\subsection{The Magic resource in the parameter space}
%%%%%%%%%%%%%%%%%%%%%%%%%%%%%%%%%%%%%%%%%%%%%%%%%%

In the following, we explore the magic resource for both the qubit and bosonic degrees of freedom in the parameter space by analyzing the ground and first excited states of the reduced system, as these low-energy states are generally easily probed and experimentally accessible. In particular, we study the $\text{mana}$, as we have seen from previous sections, it captures well the resource quantity without capturing the oscillations related to the purity of the reduced state.

In Fig.~\ref{fig:6}, we show the results as a function of the bias $\epsilon$ and coupling $g$ in resonance $\omega=2\Delta$. As a guideline, we plot the energy of the ground and first excited states in Fig.~\ref{fig:6} (a) and (b), respectively, where we obtain the standard results where, for $\epsilon=0$, energy goes from a constant value, then decreases for increasing coupling, signaling a strongly light-matter correlated state.

In Fig.~\ref{fig:6}, we observe the mana for the qubit-reduced system for the ground (c) and first excited state (d). Our reference value is the mana of $H$-type states, $\text{mana}_{H}=0.271553$, the most magical state available in the region of the Bloch sphere to which the system is restricted (the plane $z-x$.
For the ground state, we observe that for $\epsilon=0$, non-stabilizerness is minimal and equal to zero. However, increasing or decreasing it towards $\epsilon/\omega=\pm0.5$ increases it. As we discussed above, this can be understood in terms of a rotation of the qubit in the Bloch sphere around the $x$-axis, produced by the asymmetric term in the Hamiltonian. Mana almost reaches the value of the $H$ state, but increasing the coupling toward the USC will cause mana to decrease quickly. As discussed previously, light-matter interaction competes with the effect of asymmetry, canceling it out. However, one still gets around $54\%$ of the $H$ state's mana in the USC regime.

An interesting effect arises for the first excited state. For $2\epsilon=m\omega$ with $m=-1,0,1$, $\text{mana}$ goes to zero as a function of the light-matter coupling. It seems that the first indication that a hidden symmetry in the AQRM is manifested at $\epsilon/\omega=\pm0.5$. Then, a fork-like shape emerges where $\text{mana}$ goes to zero. For larger values of $\epsilon/\omega$, one recovers values similar to those of the ground state. There is evidence of suppression of non-stabilizerness in higher excited states for values $2\epsilon=\pm n\omega$, with $n\in\mathbb{Z}$. Our numerical inspection suggests that the $k$-th excited state seems to have its non-stabilizerness suppressed for the values $n=-k,-(k-1),\ _{\cdots}\ ,k-1,k$.

In Fig.~\ref{fig:6} (e) and (f), we plot $\text{mana}_{\text{bos}}$, i.e., the Wigner logarithmic negativity, for the boson-reduced ground and first excited states as well. Our reference bosonic mana is that of a Fock state with a single photon ($\text{mana}_{\text{bos}}^{(n=1)}=0.512$) because it is the nearest state in number of bosons. Following the results in Fig.~\ref{fig:5}, the ground state has zero mana in all the parameter space, which is consistent with its zero-boson nature. Conversely, the bosonic $\text{mana}$ of the first-excited state shows an increase for small light-matter coupling with the asymmetry, whose value is around the one-photon Fock state, as we have already observed in Fig.~\ref{fig:5}. The effect of the asymmetry term on the bosonic reduced density matrices is the appearance of nondiagonal terms, which arise from effective pseudospin-conserving transitions enabled by $\hat{\sigma}_x$. Hence, the results for positive and negative $\epsilon$ are not symmetric because changing the sign of $\epsilon$ implies a change in the sign of the non-diagonal coefficients. 

Also, notice that in the case of the first excited state, selecting bias values just above $\epsilon/\omega=\pm0.5$ leads to regimes where the magic resource appears for both degrees of freedom; however, care must be taken because this resource in the qubit subspace is lost for large $\abs{\epsilon}$. In general, there is no relation between the discrete and continuous resources. 

In the search for a high-magic state in the USC light-matter coupling regime, in Fig.~\ref{fig:7}, we explore the same quantities now as a function of the light-matter detuning $\delta=\omega-2\Delta$. Changing the detuning redistributes the light and matter content of the individual eigenstate, so shifting it to larger values, either negative or positive, decouples the system. Again, in Fig.~\ref{fig:7} (a) and (b), we show the ground and first excited state, respectively, in the USC regime ($g/\omega=1$), as a function of $\delta$. In Fig.~\ref{fig:7} (c), we observe that for $\epsilon=0$ and $\delta=1$, the magic resource is negligible, as one gets a stabilizer state. We recall that $2\Delta=(\omega-\delta)$, so at $\delta/\omega=1$, the $\hat{Z}$ in Eq.~\ref{eq:AQRM_Ham} cancels out, leading to a stabilizer state, but from the $\hat{X}$ operator. Again, the asymmetry is responsible for introducing non-stabilizerness in the reduced system, as shown in Figs.~\ref{fig:7} (c) and (d). 

Insights on how to explain the peculiar shape of Fig. \ref{fig:7}~(c) can be obtained while studying the non-interacting Hamiltonian in Eq.~\ref{eq:AQRM_Ham} ($g=0$). Its eigenstates are just $\ket{n}\otimes\ket{z',\pm}$, where $\ket{z',\pm}$ are the eigenstates of a new $\hat{Z}$ operator rotated an angle $\phi$ in the $z-x$ plane,  defined as $\tan(\theta)=\epsilon/\Delta$. The position of the atomic reduced density matrix becomes one of the four $\ket{H}$ states of the $z-x$ plane, if one imposes the condition $\tan(\theta)=\pm1$. Then, one obtains the following relation $2\epsilon = \pm(\omega-\delta)$ in the $\epsilon-\delta$ parameter space, a geometric place with the form of an X with its center around the point $(\omega,0)$. The maximum value of the resource is then attained in the diagonals centered at $(1,0)$ in Fig. \ref{fig:7}~(c). Similar phenomenology is obtained for the $g\neq0$ cases, but the H state is not reached.\par

At $\delta/\omega=-1$, in the USC regime, the largest value is $\sim 77\%$ of the $|H\rangle$ state's mana. A more complicated structure emerges in the case of the first excited state as shown in Fig.~\ref{fig:7} (d), so in that regime, mana goes to zero. Also, as observed in Fig.~\ref{fig:7} (e), getting out-of-resonance does not introduce non-stabilizerness to the ground state, but it does to the first excited one. In agreement with what we observed as a function of the light-matter coupling, the Wigner logarithmic negativity is close to that of a single photon state in the case of the first excited state [see Fig.~\ref{fig:7} (f)]. Also, going to negative detuning increases the bosonic $\text{mana}$.

Our results demonstrate that even in USC, one can prepare states that are close to the $|H\rangle$ states by manipulating the bias and detuning. 

%%%%%%%%%%%%%%%%%%%%%%%%%%%%%%%%%%%%%%%%%%%%%%%%%%
%%%%%%%%%%%%%%%%%%%%%%%%%%%%%%%%%%%%%%%%%%%%%%%%%%
\section{Discussion and Conclusions}
\label{sec:6}

In this work, we have quantitatively explored the non-stabilizerness and the bosonic magic resource in the reduced qubit and boson reduced subsystems of individual eigenstates in the Jaynes-Cummings and Asymmetric Quantum Rabi Models as a function of their parameters, namely, the light-matter coupling, bias, and detuning. We have identified relevant values of the parameters to induce the magic resource in the spectrum, and discussed its behavior using the Dai-Fu-Luo witness and mana monotone. For the non-stabilizerness in the qubit reduced subspace, the interaction between the qubit and the boson degree of freedom produces quantum correlations. As a general result, the bias value producing the asymmetry introduces a rotation in the Bloch sphere that leads to magic states, but the light-matter interactions suppress them. This can be understood within the polaron frame, where it is clear that in the USC regime ($\Delta/g\ll 1$) light-matter coupling favors the effect of the asymmetry, leading to a stabilizer state from the $\hat{X}$ operator.

In the weak coupling regime, we found that the maximum of non-stabilizerness for the ground state is obtained around $\epsilon=\pm\omega/2$, a value of interest because it restores a hidden symmetry in the AQRM, making the reduced eigenstate approach a $|H\rangle$ state —a state with high magic resource. Hence, the weak coupling regime is adequate for preparing this resource. Our results indicate that light-matter interaction decreases the amount of the magic resource. However, the resource is still prevalent in the USC regime. Also, a fairly good approximation of an $|H\rangle$ state can be obtained in the ground state by changing the asymmetry term and the detuning. There is evidence that non-stabilizerness is sensitive to the restoration of the symmetry of the AQRM as it vanishes in the excited states for values of the form $2\epsilon=p\omega$ with $p\in\mathbb{Z}$.

We also studied the Wigner logarithmic negativity for the boson-reduced state. We discovered that, even though the ground state is always non-magical, in the sense that it is always Gaussian, the excited states exhibit significant amounts of the bosonic magic resource, which oscillate as one goes higher in energy; however, they are related to the average photon number described by the state. As the Fock basis is the exact solution in the uncoupled case, there is bosonic magic resource before involving the light-matter interaction. 

Given that hybrid light-matter systems allow for combined degrees of freedom, they pose an attractive tool to engineer quantum resources in hybrid systems with discrete and continuous degrees of freedom~\cite{Wallquist2009,Kurizki2015}, in particular, to protect correlations from external noise and decoherence~\cite{Biswas2007}. The individual eigenstates of the AQRM Hamiltonian are light-matter entangled states that can be exploited as non-classical resources~\cite{Wang2018}, for example, via quantum control~\cite{Joshi2016}. Regarding the qubit subspace, it will be plausible to implement another bias term proportional to $\sigma_{y}$ in the Hamiltonian Eq.~(\ref{eq:AQRM_Ham}) to try reaching the most magical states in the complete Bloch sphere, the well-known T-type states~\cite{Bravyi2005}, nevertheless, this addition would increase the complexity of the problem. Refined magic states may be generated with level engineering protocols on the ground state~\cite{Liu2019}. 

Our work contributes to describing non-stabilizerness outside the context of quantum computation in relevant physical systems. Although there are other protocols for preparing magic states~\cite{Howard2017}, such as exploiting Kerr-like interactions~\cite{Boudreault2025arXiv}, we present a benchmark that can be implemented in hybrid systems to leverage features in the USC regime, which lies within the parameter space studied, where the quantum light-matter states exhibit a faster response.\\

%%%%%%%%%%%%%%%%%%%%%%%%%%%%%%%%%%%%%%%%%%%%%%%%%%
%%%%%%%%%%%%%%%%%%%%%%%%%%%%%%%%%%%%%%%%%%%%%%%%%%
\section*{Acknowledgements} 
We thank S. Pilatowski-Cameo for his insight on efficiently calculating Wigner functions. ACU thanks V. H. T. Brauer for helpful discussions related to working in the Bloch sphere and M. Cerezo for enlightening discussions about the quantifiers of nonstabilizerness. ACU acknowledges the National Postgraduate Studies Scholarship from CONAHCyT/Secihti. EBR and AEPR acknowledge EPM (CONAHCyT/Secihti). MABM acknowledges the financial support from CONAHCYT/Secihti No. CBF2023-2024-1765, the PIPAIR 2024 project from the DAI-UAM, and the Marcos Moshinsky Fellowship.
%%%%%%%%%%%%%%%%%%%%%%%%%%%%%%%%%%%%%%%%%%%%%%%%%%
%%%%%%%%%%%%%%%%%%%%%%%%%%%%%%%%%%%%%%%%%%%%%%%%%%

\appendix

%%%%%%%%%%%%%%%%%%%%%%%%%%%%%%%%%%%%%%%%%%%%%%%%%%
\section{Discrete to Continuous Heisenberg-Weyl groups}
\label{app:1}
%%%%%%%%%%%%%%%%%%%%%%%%%%%%%%%%%%%%%%%%%%%%%%%%%%

A qudit is a $d$-level system spanned in a Hilbert space $\mathcal{H}_{d}$ given by a set of states $\{|k\rangle\}$, where the $k=1,..,d$ and obeys modulo $d$ arithmetic~\cite{Axler2023}. Each normalized basis element $\ket{k}$ can be represented as a column vector with all coefficients being zero except for the $(k+1)$-th one. We refer to this set as the {\it computational basis}. 

In addition to the generalized Pauli or Heisenberg-Weyl matrices from the main text~\cite{Weyl1927}, employing other bases to represent the qudit and its operations is possible. For example, the generalized Gell-Mann matrices~\cite{Hioe1981}, a set of $d$ matrices consisting of $d(d-1)/2$ symmetric $g^{\text{S}}$, $d(d-1)/2$ asymmetric $g_{d}^{\text{A}}$, and $(n-1)$ diagonal $g_{d}^{\text{D}}$ matrices, whose matrix elements are given by
\begin{gather}
[\hat{g}_{d}^{\text{S}}]_{jk}=\sqrt{\frac{d}{2}}\left(\ketbra{k}{j}+\ketbra{j}{k}\right) \\ 
[\hat{g}_{d}^{\text{A}}]_{jk} =-i\sqrt{\frac{d}{2}}\left(\ketbra{k}{j}-\ketbra{j}{k}\right),
\end{gather}
and
\begin{gather}
\hat{g}_{d,l}^{\text{D}}=\sqrt{\frac{d}{l(l+1)}}\left(\sum_{j=1}^{l}\ketbra{j}-l\ketbra{l+1}{l+1}\right),
\end{gather}
where $1\leq j<k\leq d$ and $1\leq l \leq d-1$. In particular, for $d=2$ we recover the standard Pauli matrices $\hat{X}=\hat{g}_{2}^{\text{S}}$ and $\hat{Y}=\hat{g}_{2}^{\text{A}}$ (shift operators), and $\hat{Z}=\hat{g}_{2,1}^{\text{D}}$ (phase or clock operator), and the identity $\hat{\mathbb{I}}_{2}=\hat{g}_{2,2}^{\text{D}}$.

Another suitable choice is the eigenstates of the angular momentum operators $(\hat{J}^2,\hat{J}_z,\hat{J}_{\pm})$ that follow the $\text{su(2)}$ algebra, i.e., the Dicke states $|j,m\rangle=1/(j+m)!\binom{2j}{j+m}^{-\frac{1}{2}}\left(\hat{J}_+\right)^{j+m}|j,-j\rangle$, with $\hat{J}_{-}\ket{j,-j}=0$, and $j=(d-1)/2$, so one has $d=2j+1$ states. The corresponding change of basis is
\begin{gather}
    \braket{k}{j,m}= \frac{1}{(j+m)!}\binom{2j}{j+m}^{-\frac{1}{2}}\mel{0}{\left(\hat{X}^{\dagger}\right)^k\left(\hat{J}_+\right)^{j+m}}{0},
\end{gather}
given that $\ket{0}=\ket{j,-j}$.

%%%%%%%%%%%%%%%%%%%%%%%%%%%%%%%%%%%%%%%%%%%%%%%%%%
\subsection{Limit of the discrete Heisenberg-Weyl algebra to the continuous one}
\label{app:limitCont}
%%%%%%%%%%%%%%%%%%%%%%%%%%%%%%%%%%%%%%%%%%%%%%%%%%

The discrete Heisenberg-Weyl algebra can be extended to the infinite-dimensional limit. To this end, we define the phase and shift operators in Eqs.~(\ref{eq:z}) and~(\ref{eq:x}) in terms of a set of discrete, renormalized momentum $\hat{p}$ and position $\hat{q}$ operators that perform translations over the grid of states in the computational basis as $\hat{Z}=\exp(\sqrt{2\pi/d}\,i\hat{q})$ and $\hat{X}=\exp(-\sqrt{2\pi/d}\, i\hat{p})$~\cite{Asadian2016}. To tend to the infinite-dimensional limit $d\rightarrow\infty$, we write the discrete displacement operators in terms of a complex displacement amplitude $\alpha_{k,l}=x+ip$, where $x=k\sqrt{2\pi/d}$ and $p=l\sqrt{2\pi/d}$ are renormalized discrete displacements over the grid of states. Using this in Eq.~(\ref{eq:d}) one gets
\begin{gather}
\hat{D}(p,x)=e^{-i\frac{xp}{2}}e^{-ix\hat{p}}e^{ip\hat{x}}.
\end{gather}
In the limit of $d\rightarrow\infty$, $\alpha_{k,l}\rightarrow\alpha$ tends to the continuous, and using the Baker-Campbell-Hausdorff formula, we obtain the standard displacement operators in Eq.~(\ref{eq:dinf}). This strengthens the correspondence between the resource theory, as expressed in terms of the negativity of the Wigner function, for both discrete and continuous degrees of freedom, except that in the continuous case, the Hilbert space becomes unbounded.

%%%%%%%%%%%%%%%%%%%%%%%%%%%%%%%%%%%%%%%%%%%%%%%%%%
\section{Numerical method}
\label{app:numericalmethod}
%%%%%%%%%%%%%%%%%%%%%%%%%%%%%%%%%%%%%%%%%%%%%%%%%%

To obtain the spectrum from the main text, we solve numerically the AQRM by diagonalizing its matrix representation in the tensor product of Fock states $|n\rangle$ and spin states  $|s\rangle=|j=1/2,s\rangle$, with $s=\pm1/2$. Given that the bosonic subspace is infinite-dimensional, a truncation $n_{\text{max}}$ must be applied, large enough to avoid convergence problems. In Fock basis, the AQRM eigenstate with eigenergy $E_{k}$ is given by
\begin{align}
    \ket{E_k} &= \sum_{n=0}^{n_{\text{max}}}\sum_{m_{s}=-1/2}^{1/2} C_{n,m_z}^{(k)}\ket{n}\otimes\ket{1/2,m_{s}},
\end{align}
where $C_{n,m_{s}}^{(k)}=\langle n;1/2,m_{s}|E_{k}\rangle$ are its amplitudes in the Fock basis. Truncation implies a loss of information, so a convergence criterion must be applied to the eigenstates to be sure they are suitable for posterior studies. The chosen convergence criterion is
\begin{align}
    \sum_{n}\abs{C_{n}^{(k)}}^2\leq10^{-6}.
\end{align}
In the chosen basis, the pure density matrix generated with an eigenstate is given by
\begin{gather}
    \hat{\rho}_{k} =\ketbra{E_{k}}
    =\sum_{n,n'=0}^{n_{\text{max}}}\ \sum_{s,s'=-1/2}^{1/2}C_{n,s}^{(k)}\bar{C}_{n',s'}^{(k)}\ketbra{n,s}{n',s'},
\end{gather}
where $C_{n,s}^{k}=\langle E_{k}|n,s\rangle$. Tracing over the bosonic subspace $\text{B}$, we get
\begin{gather}
    \hat{\rho}_{k,\text{S}}= \tr_{\text{B}}\left(\hat{\rho}_{k}\right)=\\ \nonumber
    \sum_{n''=0}^{n_{\text{max}}}\mel{n''}{\hat{\rho}_{k}}{n''}
=\sum_{s,s'=-1/2}^{1/2}\left(\sum_{n''=0}^{n_{\text{max}}}C_{n'',m}^{(k)}\bar{C}_{n'',m'}^{(k)}\right)\ketbra{s}{s'},
\end{gather} 
and tracing over the atomic subspace $\text{S}$, we obtain
\begin{gather}
    \hat{\rho}_{k,\text{B}} = \tr_{\text{S}}\left(\hat{\rho}_{k}\right)=\sum_{s''=-1/2}^{1/2}\mel{s''}{\hat{\rho}_{k}}{s''}  \\ \nonumber
=\sum_{n,n'=0}^{n_{\text{max}}}\left(\sum_{s''=-1/2}^{1/2}C_{n,s''}^{(k)}\bar{C}_{n',s''}^{(k)}\right)\ketbra{n}{n'}.
\end{gather}

%%%%%%%%%%%%%%%%%%%%%%%%%%%%%%%%%%%%%%%%%%%%%%%%%%
\subsection{Bosonic Wigner function calculation}
%%%%%%%%%%%%%%%%%%%%%%%%%%%%%%%%%%%%%%%%%%%%%%%%%%

In general, a density matrix that belongs to the continuous light subspace can be expressed as
\begin{align}
  \hat{\rho}_{\text{B}} &= \sum_{n,n'}C_{n,n'}\ketbra{n}{n'}.
\end{align}
To compute the bosonic Wigner function, we exploit its linearity with the density matrices,
\begin{align}
    \mathcal{W}_{\hat{\rho}}(q,p) &= \sum_{n,n'}C_{n,n'}\mathcal{W}_{\ketbra{n}{n'}}(q,p),
\end{align}
where the coefficients $C_{n,n'}$ can be obtained analytically or by numerical means. Then, the Weyl transform of the transition operators $\ketbra{n}{n'}$ has the analytical expression
\begin{align}
    \mathcal{W}_{\ketbra{n}{n'}}(r,\phi) &= \notag\\
    & \hspace*{-1.5cm}\begin{cases}
        \displaystyle \frac{(-1)^n}{4\pi}\sqrt{\frac{n!}{n'!}}\ \left(2r^2\right)^{\frac{\abs{n'-n}}{2}}\text{e}^{-r^2}L_{n}^{\abs{n'-n}}(2r^2)\cdot\text{e}^{-i\abs{n'-n}\phi}\qquad\\
        \hspace*{6.5cm}\text{for }n\leq n'\\[0.35cm]
        \displaystyle \frac{(-1)^{n'}}{4\pi}\sqrt{\frac{n'!}{n!}}\ \left(2r^2\right)^{\frac{\abs{n-n'}}{2}}\text{e}^{-r^2}L_{n'}^{\abs{n-n'}}(2r^2)\cdot\text{e}^{i\abs{n-n'}\phi}\\
        \hspace*{6.5cm}\text{for }n> n'.
    \end{cases} 
\end{align}
The Wigner function is computed inside the finite region where it is non-vanishing, and then it is renormalized.

%%%%%%%%%%%%%%%%%%%%%%%%%%%%%%%%%%%%%%%%%%%%%%%%%
%%%%%%%%%%%%%%%%%%%%%%%%%%%%%%%%%%%%%%%%%%%%%%%%%
\section{2-dimensional discrete Wigner functions}
\label{app:DiscreteWigner}
%%%%%%%%%%%%%%%%%%%%%%%%%%%%%%%%%%%%%%%%%%%%%%%%%
%%%%%%%%%%%%%%%%%%%%%%%%%%%%%%%%%%%%%%%%%%%%%%%%%

Discussions about a general definition of a discrete Wigner function for all dimensions arose since its conception by Wootters~\cite{Wootters1987}, namely,
\begin{gather}
    \hat{W}^{(W)}_{k,l}= \frac{1}{2}\left[
    \hat{\mathbb{I}}_{(2)}+(-1)^{l}\hat{X}+(-1)^{k+l}\hat{Y}+(-1)^k\hat{Z}\right]
\end{gather}
is the Wootters-Wigner operator for $d=2$. In recent years there have been many proposals towards a unified definition of a discrete Wigner function~\cite{Tilma2016,Meng2023,Rundle2020,Marchiolli2019,SanchezSoto2025}. In this work, we use the one proposed by Marchiolli et al.~\cite{Marchiolli2019}:
\begin{gather}
\hat{W}^{(M)}_{k,l}=\frac{1}{2}\left[\mathbb{I}_{(2)}+(-1)^{l}\hat{X}+(-1)^{k+l+1}\hat{Y}+(-1)^{k}\hat{Z}\right].
\end{gather}
Note that due to the symmetry of the systems used in this work, all considered models do not possess any term proportional to $\hat{Y}$. Therefore, both, $\hat{W}^{(W)}_{k,l}$ and $\hat{W}^{(M)}_{k,l}$ will have the same expected values for AQRM-like systems.

%%%%%%%%%%%%%%%%%%%%%%%%%%%%%%%%%%%%%%%%%%%%%%%%%%
\section{Characteristic function and von Neumann entropy relation}
\label{app:conjecture}

The Bloch vector of a two-level system is given by
\begin{gather}
	\hat{\rho} = \mqty(a & c \\ \bar{c} & b)=\frac{1}{2}\left(\hat{I}+\vb{s}\cdot\hat{\vb{\sigma}}\right),
	\label{eq:densmat}
\end{gather} 
with the Bloch vector $\vb{s} = \big(s_{x},s_{y},s_{z}\big)=\big(2\Re(c),\ -2\Im(c),\ a-b\big)$ with $a+b=1$. In spherical coordinates, useful to the symmetries discussed in this work, the Bloch vector becomes $\vb{s} = (\abs{\vb{s}},\theta,\varphi)$, where $\abs{\vb{s}} = \sqrt{(a-b)^2+4\abs{c}^2}$, $\cos(\theta) = (a-b)(\sqrt{(a-b)^2+4\abs{c}^2})$, and $\tan(\varphi) = -\Im(c)/\Re(c)$.

In terms of the Bloch vector in spherical coordinates, the von Neumann entropy $S(\hat{\rho}) = -\tr(\hat{\rho}\ln(\hat{\rho}))$ and Dai-Fu-Luo's witness in Eq.~(\ref{eq:DaiFuLuo}) are expressed as
\begin{gather}
	S(\abs{\vb{s}}) = \ln(2)-\frac{1}{2}\ln[
	(1-\abs{\vb{s}})^{(1-\abs{\vb{s}})}
	(1+\abs{\vb{s}})^{(1+\abs{\vb{s}})}
	],\\
  \mathcal{M}(\vb{s}) = 1+\abs{\vb{s}}\abs{\sin(\theta)}\left[\abs{\cos(\phi)}+\abs{\sin(\phi)}+\abs{\cot(\theta)}\right]
\end{gather}
The derivative of the von Neumann entropy with respect to $\abs{\vb{s}}$ is given by
\begin{gather}
  \dv{S}{\abs{\vb{s}}} = -\text{arctanh}(\abs{\vb{s}})=-\abs{\vb{v}}+\mathcal{O}(\abs{\vb{v}}^3).
\end{gather}
Here, the reduced density matrices of the eigenstates of Hamiltonians with parity symmetry are diagonal. Hence, $\Re(c)=\Im(c)=0$ and $\cos(\theta)=\pm1$ depending on the sign of $a-b$ or, equivalently, $\sin(\theta)=0$. This yields $\mathcal{M}(\vb{s})=1+\abs{\vb{s}}$, or
\begin{gather}
  \mathcal{M}(\vb{s}) = 1-\dv{S}{\abs{\vb{s}}}-\mathcal{O}(\abs{\vb{s}}^3)
\end{gather}
This result is valid for parity-conserving Hamiltonians, and once the asymmetry term $\epsilon$ is present, the previous condition no longer holds. Nonetheless, from Figs.~\ref{fig:1},~\ref{fig:3}, and~\ref{fig:4}, it can be noticed by numerical inspection that the von Neumann entropy and $\mathcal{M}$ still reflect each other, so we conjecture that it holds $\mathcal{M} \propto -\partial S/\partial E$, where $E$ is the energy of the system. However, we have not proved this conjecture. 

%%%%%%%%%%%%%%%%%%%%%%%%%%%%%%%%%%%%%%%%%%%%%%%%%%

\bibliography{references}

\end{document}